\documentclass{jpsj-suppl}
\usepackage{txfonts} 

\title{Study of Baryon Resonances in the Photoproduction $\gamma p 
\rightarrow K^*\Sigma$(1190)}

\author{Sang-Ho \textsc{Kim}$^{1}$, Atsushi \textsc{Hosaka}$^{1}$, 
Seung-il \textsc{Nam}$^{2}$ and Hyun-Chul \textsc{Kim}$^{3}$}

\inst{$^{1}$ Research Center for Nuclear Physics (RCNP), Osaka 567-0047, 
Japan  \\
$^{2}$School of Physics, Korea Institute for Advanced Study (KIAS), 
Seoul 130-722, Republic of Korea \\
$^{3}$Department of Physics, Inha University, Incheon 402-751, 
Republic of Korea}

\email{shkim@rcnp.osaka-u.ac.jp, hchkim@inha.ac.kr}

\recdate{July 15, 2013}

\abst{In this talk, we report the theoretical studies of baryon resonances for 
the $\gamma p \rightarrow K^{*0}\Sigma^+$ and $\gamma p \rightarrow
K^{*+}\Sigma^0$ reaction processes by making use of the effective
Lagrangian method in the tree-level Born approximation. The numerical
results for the total and differential cross sections are shown with
the presently available data. It turns out that the contributions from
the $N^*$ and $\Delta^*$ resonances are almost negligible near the
threshold energy. On the other hand, the effects of the $K$  
exchange in the $t$ channel and $\Delta(1232)$ in the $s$ channel 
are considerably larger than other ones in the scattering process,
being in good agreement with the experimental data}

\kword{$K^*\Sigma$ photoproduction, effective Lagrangian, $N^*$ and
  $\Delta^*$ resonances} 

\begin{document}
\maketitle

\section{Introduction}
Various efforts to understand the baryon resonances and to find {\it
  missing} resonances have been made for decades via the hadron
produtcion processes. Photoproduction is one of them, and,
especially, it tuns out that the strange meson-baryon production off
the nucleon target is very useful. In the present talk, we would like
to investigate the reaction process of  $K^{*}\Sigma(1190)$
photoproduction off the proton target. Recently, CBELSA/TAPS
collaboration at Electron Stretcher and Accelerator
(ELSA)~\cite{Nanova:2008}, the CLAS collaboration at Thomas Jefferson  
National Accelerator Facility (Jefferson
Lab)~\cite{Hleiqawi:2007,Wei:2013}, and LEPS collaboration at Super
Photon Ring-8 GeV (Spring-8)~\cite{Hwang:2012} have provided the
experimental data for this reaction channel. The related theoretical
studies also exist in both models using the chiral quark
model~\cite{Zhao:2001} and effective Lagrangian
method~\cite{Oh:20006}. The theoretical prediction from
Ref.~\cite{Zhao:2001} turned out to be hard to reproduce the recent
experimental data. In Ref.~\cite{Oh:20006}, the effective Lagrangian
method was employed at the tree-level Born approximation,  
considering the $K$ and $\kappa$ exchanges in the $t$ channel, $N$-
and $\Delta(1232)$-pole contributions in the $s$ channel, and hyperons
($\Sigma, \Sigma^*$) in the $u$ channel. The motivation of our
present work is to include all the possible $N^*$ and $\Delta^*$
resonances, in a full relativistic manner, such as $D_{13}(2080)$,
$S_{11}(2090)$, $G_{17}(2190)$, $D_{15}(2200)$, $S_{31}(2150)$,
$G_{37}(2200)$, and  $F_{37}(2390)$.   
\section{Formalism}
We employ the effective Lagrangian method in the tree-level Born
approximation. The relevant diagrams are shown in Fig.\ref{figures:1},
and the effective Lagrangians for the interaction vertices are defined
in Ref.~\cite{SHKim1:2012}. The coupling constants are obtained by
using experimental data~\cite{Nakamura:2010} and the SU(6)
relativistic quark model~\cite{Capstick1:1992,Capstick2:1998}. 
As for the resonances,
we use the following relation:    
\begin{figure}[t]
\vspace{-1.7em}
\begin{center}
\includegraphics[width=13cm]{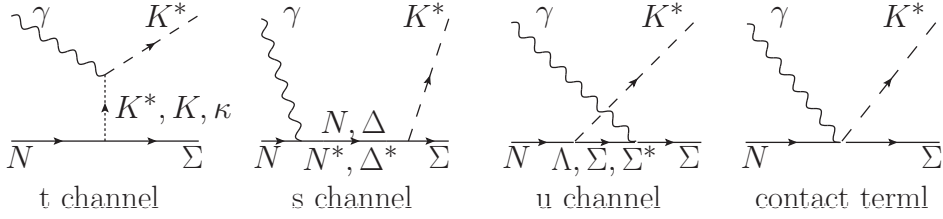}
\caption{Relevant tree-level Feynman diagrams for $\gamma p \to
  K^{*}\Sigma$.}      
\end{center}
\label{figures:1}
\end{figure}
\begin{eqnarray}
\label{eq:GAMMA1}
\Gamma(R \rightarrow
N\gamma)=\frac{k_\gamma^2}{\pi}\frac{2M_N}{(2j+1)M_R} 
[|A_{1/2}|^2+|A_{3/2}|^2],\,\,\;\;\;
\Gamma(R\to K^*\Sigma)=\sum_{l,s}|G(l,s)|^2,
\end{eqnarray}
for the photo- and strong-coupling constants, respectively. Here, $A_{1/2}$
and $A_{3/2}$ represent the helicity amplitudes, computed by using the 
experimental or theoretical values~\cite{Capstick1:1992,Nakamura:2010}, from 
which the transition magnetic moments are given~\cite{SHKim1:2012}. In addition, 
the scattering amplitudes $G(l,s)$ are estimated by the SU(6) quark 
model~\cite{Capstick2:1998}, based on which the strong coupling
constants are calculated~\cite{SHKim1:2012}. For simplicity, we
consider only the resonance states $\sqrt{s}\lesssim2500$ MeV, since 
we are interested in the vicinity of the threshold. The scattering
amplitudes can be written with the phenomenological form factors  
that satisfy the Ward-Takahashi identity as follows: 
\begin{eqnarray}
\label{eq:AMP}
\mathcal{M}(\gamma p \to K^{*0}\Sigma^+)&=&
[\mathcal{M}_{s(N)}^{\mathrm{elec}}+\mathcal{M}_{u(\Sigma)}]F_\mathrm{com}^2 
+ \mathcal{M}_{s(N)}^{\mathrm{mag}} F_N^2
+ \mathcal{M}_{t(K)}F_K^2 + \mathcal{M}_{t(\kappa)} F_\kappa^2 
\cr &&
+ \mathcal{M}_{s(\Delta)} F_\Delta^2
+ \mathcal{M}_{u(\Sigma^*)} F_{\Sigma^*}^2
+ \mathcal{M}_{s(R)}F_{R}^2,
\cr 
\mathcal{M}(\gamma p \to K^{*+}\Sigma^0)&=&
[\mathcal{M}_{t(K^*)} +
\mathcal{M}_{s(N)}^{\mathrm{elec}} + \mathcal{M}_{\mathrm{c}}]
F_{\mathrm{com}}^2   
+ \mathcal{M}_{s(N)}^{\mathrm{mag}} F_N^2
+ \mathcal{M}_{t(K)}F_K^2 + \mathcal{M}_{t(\kappa)} F_\kappa^2 
\cr &&
+ \mathcal{M}_{s(\Delta)} F_\Delta^2
+ \mathcal{M}_{u(\Lambda)} F_\Lambda^2
+ \mathcal{M}_{u(\Sigma)} F_\Sigma^2
+ \mathcal{M}_{u(\Sigma^*)} F_{\Sigma^*}^2
+ \mathcal{M}_{s(R)}F_{R}^2.
\end{eqnarray}
Here, the form factors are defined generically as
\begin{equation}
\label{eq:FF}
F_\mathrm{com}
=F_{N(K^*)}F_{\Sigma(N)}-F_{N(K^*)}-F_{\Sigma(N)}, 
\qquad
F_\Phi=\frac{\Lambda^2_{\Phi}-M^2_\Phi}{\Lambda^2_\Phi-p^2},
\qquad
F_B=\frac{\Lambda^4_B}{\Lambda^4_B+(p^2-M^2_B)^2},
\end{equation}
where $p$, $\Lambda_\Phi$, and $\Lambda_B$ stand for the momentum
transfer, the cutoff masses for the meson-exchange and baryon-pole
diagrams, respectively.  The corresponding cutoff masses are
determined phenomenologically to reproduce the experimental data: 
\begin{table}[h]
\begin{center}
\begin{tabular}{cc|ccc|cc} \hline\hline
\multicolumn{2}{c|}{$\Lambda_\Phi$ for $t$ channel}&
\multicolumn{3}{|c|}{$\Lambda_B$ for $s$ channel}&
\multicolumn{2}{|c}{$\Lambda_B$ for $u$ channel}\\
\hline
$\Lambda_{K^*}$&
$\Lambda_K,\Lambda_\kappa$&
$\Lambda_N,\Lambda_{\Delta}$ &
$\Lambda_{N^*}$&
$\Lambda_{\Delta^*}$&
$\Lambda_\Lambda$&
$\Lambda_\Sigma,\Lambda_{\Sigma^*}$\\
\hline
$0.80$ GeV&
$1.15$ GeV&
$1.50$ GeV&
$1.00$ GeV&
$1.00$ GeV&
$0.70$ GeV&
$0.95$ GeV\\
\hline\hline
\end{tabular}
\caption{Cutoff masses for the form factors in Eq.~(\ref{eq:FF}) for
each channel.}   
\label{TABLE:1}
\end{center}
\end{table}
\vspace{-2em}
\section{Numerical Results}
\subsection{Total Cross Section}
\begin{figure}[t]
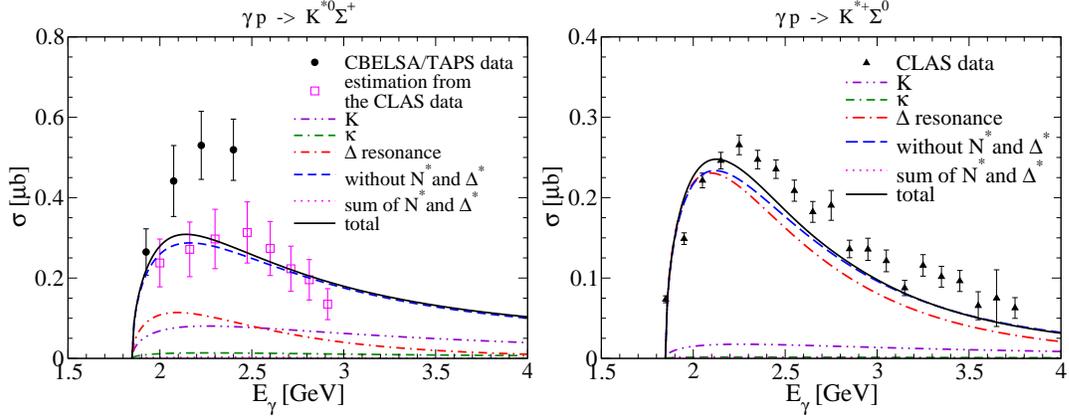

\vspace{-1.7em}
\begin{tabular}{cc}
\includegraphics[width=7cm]{FIG2a.eps}
\includegraphics[width=7cm]{FIG2b.eps}
\end{tabular}
\caption{(Color online) Total cross sections for $\gamma p \to K^{*0}\Sigma^+$ 
as functions of the photon energy $E_\gamma$ in the left panel. The experimental
data are taken from the CBELSA/TAPS~\cite{Nanova:2008} and 
CLAS~\cite{Hleiqawi:2007} collaborations. Those for $\gamma p \to K^{*+}\Sigma^0$ 
with the same notation in the right panel. The experimental data are taken from
the CLAS collaboration~\cite{Wei:2013} .}
\label{figures:TCS}
\end{figure}

The contributions of each of channels are shown
separately in Fig.~\ref{figures:TCS}. In the left panel of
Fig~\ref{figures:TCS}, it turns out that the $K$-exchange and
$\Delta(1232)$-pole contributions are crucial, and in the right panel,
it can be understood that $\Delta(1232)$-pole one almost dominates the
reaction process. From those figures, we find that the numerical
results are in good agreement with the CLAS data, and the effect of
$N^*$ and $\Delta^*$
resonance contributions turns out to be almost negligible.
\subsection{Differential Cross Section}
\begin{figure}[b]
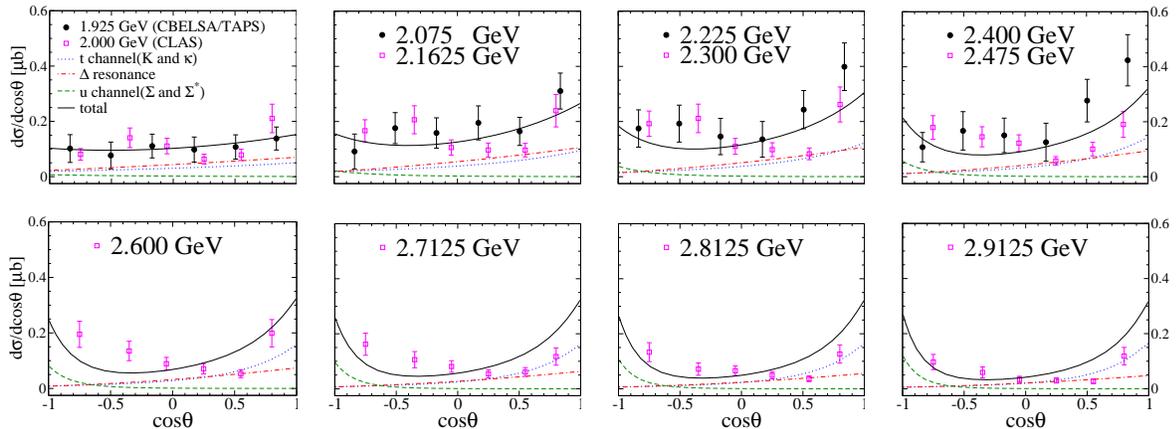

\begin{tabular}{cc}
\hspace{-1.0em}
\includegraphics[width=3.82cm]{FIG4a.eps}\hspace{1.1em}
\includegraphics[width=3.30cm]{FIG4b.eps}\hspace{1.1em}
\includegraphics[width=3.30cm]{FIG4c.eps}\hspace{1.1em}
\includegraphics[width=3.55cm]{FIG4d.eps}\vspace{0.85em}
\\ \hspace{-1.0em}
\includegraphics[width=3.87cm]{FIG4e.eps}\hspace{0.79em} 
\includegraphics[width=3.41cm]{FIG4f.eps}\hspace{0.79em}
\includegraphics[width=3.41cm]{FIG4g.eps}\hspace{0.79em} 
\includegraphics[width=3.64cm]{FIG4h.eps}
\end{tabular}
\caption{(Color online) Differential cross sections for $\gamma p \to K^{*0}\Sigma
^+$ as functions of $\cos\theta$ for different photon energies ($E_\gamma$) in the 
range $(1.925-2.9125)$ GeV. The experimental data of the CBELSA/TAPS and CLAS 
collaborations are taken from Ref.~\cite{Nanova:2008} and 
Ref.~\cite{Hleiqawi:2007}, respectively.}        
\label{figures:DCS1}
\end{figure}

\begin{figure}[t]
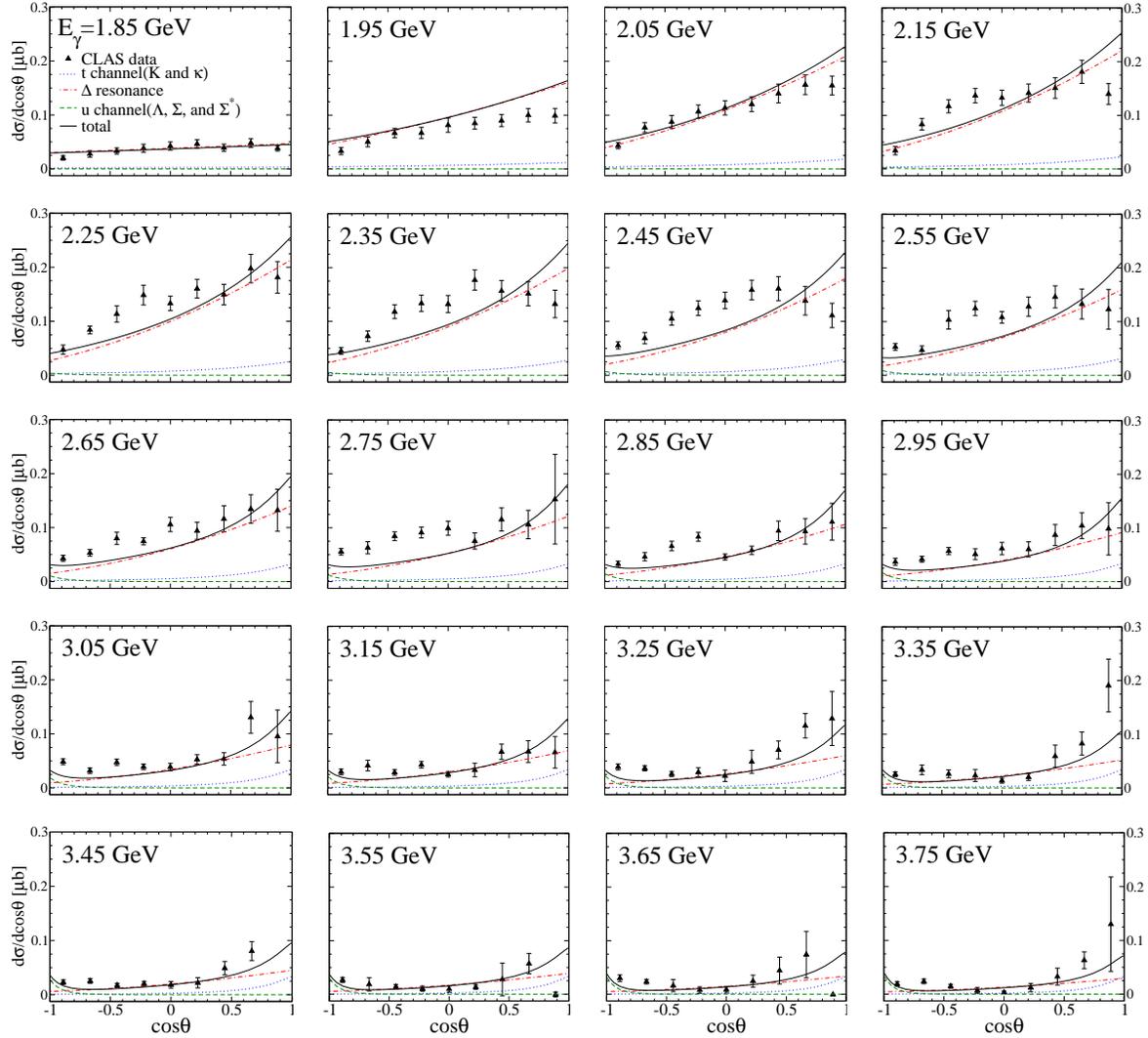

\vspace{-1.7em}
\begin{tabular}{cc}
\hspace{-1.0em}
\includegraphics[width=3.82cm]{FIG5_01.eps}\hspace{1.1em}
\includegraphics[width=3.30cm]{FIG5_02.eps}\hspace{1.1em}
\includegraphics[width=3.30cm]{FIG5_03.eps}\hspace{1.1em}
\includegraphics[width=3.55cm]{FIG5_04.eps}\vspace{0.85em}
\\ \hspace{-1.0em}
\includegraphics[width=3.82cm]{FIG5_05.eps}\hspace{1.1em} 
\includegraphics[width=3.30cm]{FIG5_06.eps}\hspace{1.1em} 
\includegraphics[width=3.30cm]{FIG5_07.eps}\hspace{1.1em} 
\includegraphics[width=3.55cm]{FIG5_08.eps}\vspace{0.85em}
\\ \hspace{-1.0em}
\includegraphics[width=3.82cm]{FIG5_09.eps}\hspace{1.1em} 
\includegraphics[width=3.30cm]{FIG5_10.eps}\hspace{1.1em} 
\includegraphics[width=3.30cm]{FIG5_11.eps}\hspace{1.1em} 
\includegraphics[width=3.55cm]{FIG5_12.eps}\vspace{0.85em}
\\ \hspace{-1.0em} 
\includegraphics[width=3.82cm]{FIG5_13.eps}\hspace{1.1em} 
\includegraphics[width=3.30cm]{FIG5_14.eps}\hspace{1.1em} 
\includegraphics[width=3.30cm]{FIG5_15.eps}\hspace{1.1em} 
\includegraphics[width=3.55cm]{FIG5_16.eps}\vspace{0.85em}
\\ \hspace{-0.9em} 
\includegraphics[width=3.87cm]{FIG5_17.eps}\hspace{0.79em} 
\includegraphics[width=3.41cm]{FIG5_18.eps}\hspace{0.79em} 
\includegraphics[width=3.41cm]{FIG5_19.eps}\hspace{0.79em} 
\includegraphics[width=3.64cm]{FIG5_20.eps}
\end{tabular}
\caption{(Color online) Differential cross sections for $\gamma p \to K^{*+}\Sigma
^0$ as functions of $\cos\theta$ for different photon energies ($E_\gamma$) in the 
range $(1.85-3.75)$ GeV. The experimental data of the CLAS collboration are 
taken from Ref.~\cite{Wei:2013}.}  
\label{figures:DCS2}
\end{figure}

Here we show the numerical results for the differential cross sections
as functions of $\cos\theta_{K^*}$ for different photon energies
$E_\gamma=(1.925-2.9125)$ and $(1.85-3.75)$ GeV for 
$K^{*0}\Sigma^+$ and $K^{*+}\Sigma^0$ production, respectively. 
As shown in Figs.~\ref{figures:DCS1}
and \ref{figures:DCS2}, we can see that the $t$-channel exchanges ($K$
and $\kappa$) play an important role in the forward angle scattering
region, wheres the backward scattering regions are dominated by the
$u$-channel contributions ($\Lambda,\Sigma,\Sigma^*$).   
\section{Summary}
We investigated the reaction processes of $\gamma p \to
K^{*0}\Sigma^+$ and $\gamma p \to K^{*+}\Sigma^0$, emphasizing on the
roles of baryon resonances. We found that the resonance contributions
gave only negligible effects. On the other hand, $K$-exchange and
$\Delta(1232)$-pole contributions turned out to be crucial for both of
reaction processes. This observation is quite different from the
$K^*\Lambda$ photoproduction~\cite{SHKim2:2011}. More details can be
found in Ref.~\cite{SHKim1:2012}. 

\end{document}